\documentclass[%
 reprint,
 amsmath,amssymb,
 aps,
]{revtex4-2}
\usepackage{graphicx}
\usepackage{float} 
\usepackage{hyperref}
\usepackage{natbib}
\usepackage{xcolor} 
\usepackage{dcolumn}
\usepackage{bm}

\usepackage[normalem]{ulem}

\begin{document}

\preprint{APS/123-QED}

\title{Magnetic field as a tuning knob for edge mode interaction in integer quantum Hall interferometers}

\author{Ning Ma$^{1}$}
\author{Tevž Lotrič$^{2}$}
\author{Kenji Watanabe$^{3}$}
\author{Takashi Taniguchi$^{4}$}
\author{Steven H. Simon$^{2}$}
\author{Mitali Banerjee$^{1,5}$}
\email{mitali.banerjee@epfl.ch}

\affiliation{$^1$ Institute of Physics, École Polytechnique Fédérale de Lausanne (EPFL), CH-1015 Lausanne, Switzerland}

\affiliation{$^2$ Rudolf Peierls Centre for Theoretical Physics, Parks Road, Oxford, OX1 3PU, UK}

\affiliation{
$^3$ Research Center for Functional Materials, National Institute for Materials Science, 1-1 Namiki, Tsukuba 305-0044, Japan}

\affiliation{
$^4$ International Center for Materials Nanoarchitectonics, National Institute for Materials Science, 1-1 Namiki, Tsukuba 305-0044, Japan}

\affiliation{
$^5$ Center for Quantum Science and Engineering (QSE Center), École Polytechnique Fédérale de Lausanne (EPFL), 1015, Lausanne, Switzerland}

\date{\today}

\begin{abstract}
\noindent 
The quantum Hall interferometer has been shown to be a powerful tool for probing both the nature of the chiral modes present at the edges of these systems and the interactions between them. Strong inter-edge mode interactions can lead the system into a regime where the period of the interferometer signal corresponds to an insertion of a half-quantum flux $h/2e$, rather than the standard flux quantum $h/e$. Using a highly tunable bilayer graphene device, we are able to measure the prevalence of both the $h/e$-period and the doubled-frequency regimes as a function of magnetic field, providing insight into the microscopic nature of the doubling mechanism. Furthermore, we are able to push our device to a regime where double-frequency oscillations coexist with a superimposed period-$h/e$ pattern associated with the inner chiral modes farther from the edge. We argue that this regime has a natural interpretation in a picture of the frequency doubling being caused by strong capacitive couplings between the edge modes.
\end{abstract}

\maketitle


\section{Introduction}\label{sec:introduction}

When confined to two dimensions and subjected to a perpendicular magnetic field, electrons exhibit the quantum Hall effect (QHE)\cite{von_klitzing_quantized_1986,von_klitzing_40_2020}, in which the electronic transport is carried out by one-dimensional chiral edge channels. These edge channels provide a versatile platform for exploring quantum coherent phenomena\cite{van_wees_observation_1989,ji_electronic_2003,de_c_chamon_two_1997}. Electrostatic manipulation of these chiral edge channels can be conducted using local gates and quantum point contacts (QPCs), enabling the electronic analogs of Fabry-Pérot\cite{van_wees_observation_1989,de_c_chamon_two_1997} or Mach-Zehnder\cite{ji_electronic_2003,law_electronic_2006,deviatov_quantum_2011} interferometers. 

Electronic Fabry-Pérot interferometry (FPI) has emerged as a powerful tool for probing the phase coherence and quasiparticle properties of QH edge channels \cite{halperin_theory_2011,nakamura_direct_2020,ronen_aharonovbohm_2021,kim_aharonovbohm_2026,samuelson_slow_2026}. In the fractional QH regime, the observation of phase jumps in the interference patterns\cite{nakamura_direct_2020,nakamura_impact_2022,kim_aharonovbohm_2024,kim_aharonov-bohm_2024,kim_selective_2026,henzinger_controlled_2026} has provided evidence for fractional quasiparticles and anyonic braiding. In the integer QH regime, an intriguing phenomenon is the emergence of $h/2e$ flux periodicity of the interfering outermost integer QH edge states\cite{choi_robust_2015,biswas_electron_2023,liang_single_2025}, in which the Aharonov-Bohm interference pattern shows an unexpected doubling of the oscillation frequency to $h/2e$, instead of $h/e$. This phenomenon emerges when multiple edge channels are involved in the FPIs and has been associated with the outer two edges belonging to the same spin-split Landau level\cite{biswas_electron_2023}.

Recently, this anomalous interference periodicity has also been observed in graphene-based FPI devices\cite{werkmeister_strongly_2024,yang_evidence_2024}. A transition from $h/e$ to $h/2e$ oscillations accompanied by phase jumps can be induced by tuning the interaction strength between edge channels through electrostatic gating\cite{werkmeister_strongly_2024} and source-drain bias voltages\cite{yang_evidence_2024}. Despite the reproducibility of this behavior in different material platforms, the mechanism behind the $h/2e$ flux period is still under debate. There are experiments in GaAs that have associated the $h/2e$ periodicity with the existence of electron pairing\cite{choi_robust_2015,biswas_electron_2023}, further supported by shot noise measurements\cite{choi_robust_2015,biswas_electron_2023} with an effective charge $e^{*}$ $\approx$ $2e$. Related experiments in graphene have reported correlated interference involving pairs and triplets of edge channels\cite{yang_evidence_2024}. An alternative interpretation attributes the phenomenon to the repulsive interaction between the co-propagating edge channels and the capacitive charging in the FPI cavity\cite{werkmeister_strongly_2024,liang_single_2025}, without requiring the formation of electron pairs or triplets. Although these observations point to a central role of inter-edge coupling, it still remains open how the coupling strength determines the interference spectrum.

Here we investigated the Aharonov-Bohm (AB) oscillations in quantum Hall Fabry-Pérot interferometry in bilayer graphene systems. By tuning independent electrostatic gates and the magnetic field to control the coupling between edge channels, we observed the doubling of frequencies in the outermost edge channel oscillations at $\nu$ = -2. In addition, we observed an oscillation frequency associated with the inner edge channel. We interpret these results within a theory of frequency doubling caused by strong capacitive coupling between the different edge modes \cite{werkmeister_strongly_2024}, which we naturally extend to explain the appearance of the inner interference channel. By varying the magnetic field, we find that both the frequency doubling and the additional frequency become more pronounced at lower magnetic fields. Analysis of the edge trajectories reveals that the evolution coincides with a reduction of the spatial separation between the inner and outer edges. Our results establish the magnetic field, together with the electrostatic gating, as an approach to controlling the inter-edge coupling and provide a picture of the interference spectrum based on the spatial separations between edge channels.

\begin{figure}
      \includegraphics[width=1\linewidth]{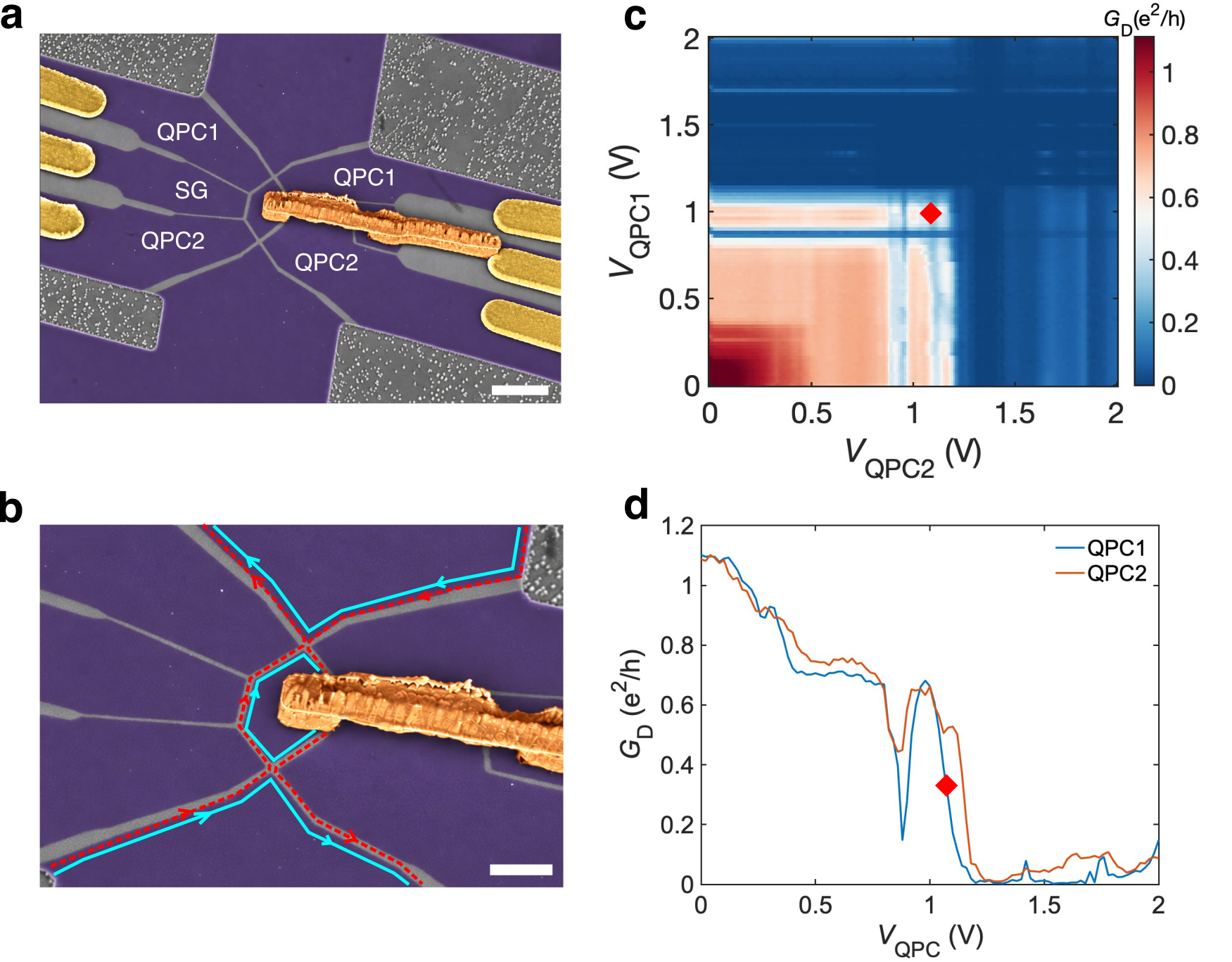}
      \caption{\textbf{Design and tunability of the interferometer at $\nu$ = -2.} \textbf{a,} False color scanning electron microscope image of the graphene interferometer device. Independent gates (purple) are labeled as the side gate (SG) and QPC pairs 1 and 2. The carrier density inside the cavity is controlled by an air bridge (orange). Scale bar: 1~$\mu m$. \textbf{b,} Magnified image of the interfering region in the center. The central gate sets the area of the interferometer to 0.4 $\mu m^{2}$. The outer and inner edge channels of $\nu$ = -2 are depicted as a red dashed line and a blue line, respectively. Scale bar: 0.5~$\mu m$. \textbf{c,} Diagonal conductance as a function of both $V_{QPC1}$ and $V_{QPC2}$ at a fixed $V_{bg}$ = -0.63 V. The red diamond shows the position of the working points of the QPCs. \textbf{d,}  Diagonal conductance as a function of either $V_{QPC1}$ or $V_{QPC2}$ at a fixed $V_{bg}$ = -0.63 V. Each curve as a function of $V_{QPC}$ is measured while the other QPC remains open. The bulk filling factor is kept at $\nu$ = -2. The diagonal conductance plateau is considerably lower than the quantum Hall conductance value, indicating the existence of significant backscattering at the QPC. The red diamond marks the position corresponding to the interfering outermost edge state at $\nu$ = -2. All the data are measured at 250 mK and 8 T.
}\label{fig:device}
      
\end{figure}

\section{Design and tuning of the interferometer at $\nu$ = -2}\label{sec:design-tuning}
We fabricated the graphene device and tuned the interferometer with several independent graphite top gates. The scanning electron microscope image is shown in Fig.~\ref{fig:device}a. The top gates were defined by a standard e-beam lithography process and oxygen plasma etching, as shown in the Methods. The two QPCs are electrostatically defined by pairs of top gates. The cavity inside the interferometer is defined by the etched top gate trenches, which are narrower than 100 $nm$. The resulting cavity area is around 0.4 $\mu m^{2}$ by design. A metallic air bridge is fabricated to independently control the cavity. 

We first characterize the bulk filling factor by tuning the bottom gate. The transmission of each QPC is independently controlled by the voltages $V_{QPC1}$ and $V_{QPC2}$ applied to the corresponding gates, shown in Fig.~\ref{fig:device}a. Figure~\ref{fig:device}d shows the diagonal conductance $G_{D}$ as a function of each QPC voltage at $\nu$ =-2 while the other QPC is kept open. The diagonal conductance $G_{D}$ decreases as the QPC voltage is increased, indicating that the edge states are pinched off. In the following, we will focus on the interference of the outermost edge at $\nu$ = -2; therefore, we have fixed both QPCs at the working points marked by the red diamond in Fig.~\ref{fig:device}c and d, corresponding to 50$\%$ transmission of the outermost edge channels at the QPCs. By tuning the voltage applied to the side gate (SG), a high-visibility oscillation can be observed in our measurements, as shown in Fig.~\ref{fig:pyama_plots_8t_nu2}. 

\section{$V_{CG}$ dependence of the oscillations at $\nu$ = -2}\label{sec:vcg-dependence}

\begin{figure*}
    \includegraphics[width=1\textwidth]{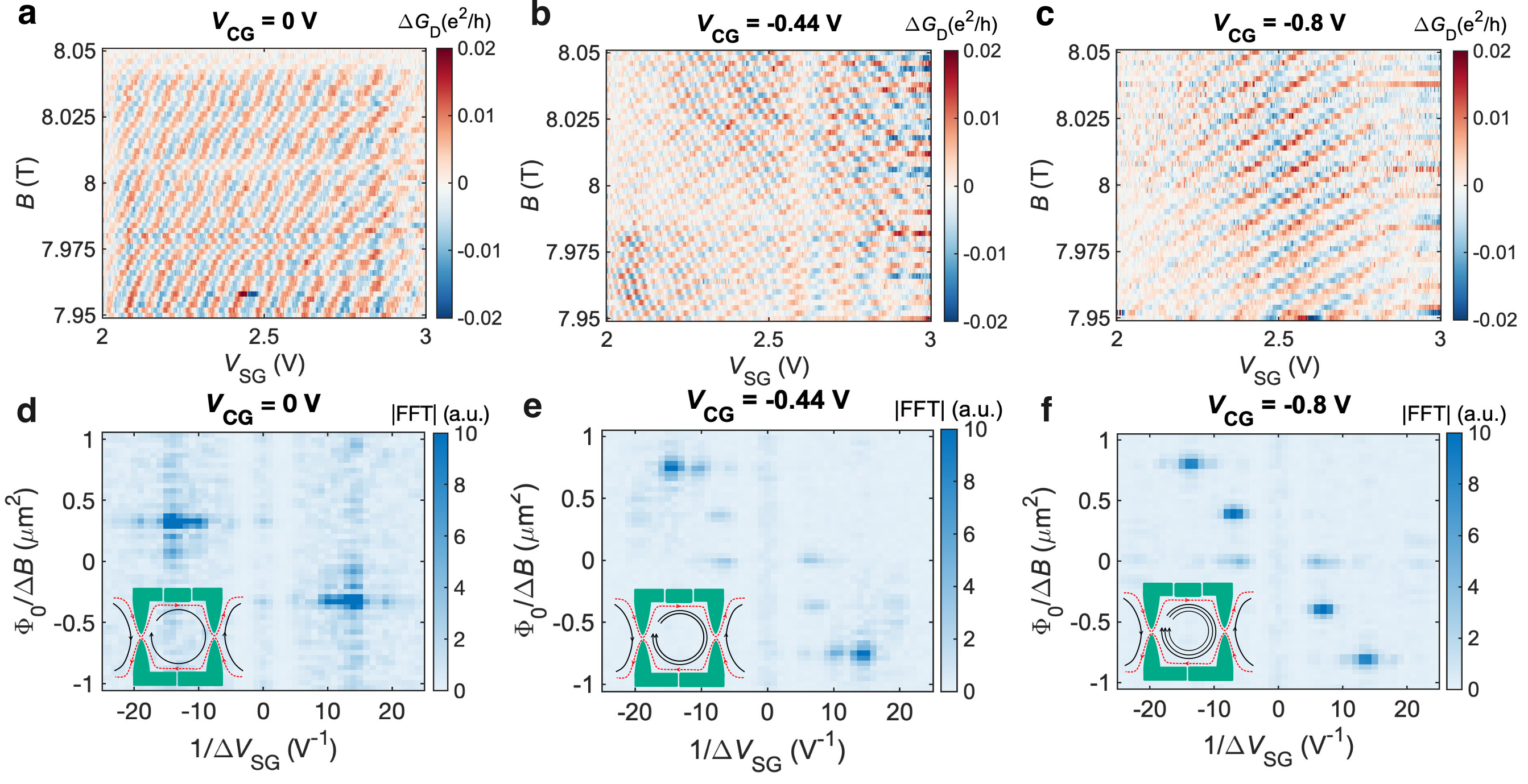}
    \caption{\textbf{Pyjama plots and the corresponding 2D FFT at bulk filling $\nu$ = -2.} \textbf{a-c,} The oscillating $\Delta G_{D}$ as a function of both side gate voltage $V_{SG}$ and magnetic field at different $V_{CG}$ values, which correspond to a continuous tuning from $\nu_{cavity}$ = -2 (\textbf{a}) to $\nu_{cavity}$ = -3 (\textbf{b}) and $\nu_{cavity}$ = -4 (\textbf{c}), respectively. The changes in the periodicity indicate that a transition from $h/e$ to $h/2e$ oscillations occurs when tuning $V_{CG}$. \textbf{d-f,} 2D FFT results of the pyjama plots. The FFT peaks also show the transition from $h/e$ to $h/2e$ periodicity from \textbf{d} to \textbf{f}. In addition, another $h/e$ frequency appears in \textbf{e} and \textbf{f} and becomes dominant with increasing $V_{CG}$. The inset schematics show the edge channel configurations at different $V_{CG}$. The red dashed lines indicate the partitioned outer edge channels and the black lines indicate the inner channels. All the data are measured at 250 mK and 8 T.
}
    \label{fig:pyama_plots_8t_nu2}
\end{figure*}

\begin{figure*}
    \includegraphics[width=0.95\textwidth]{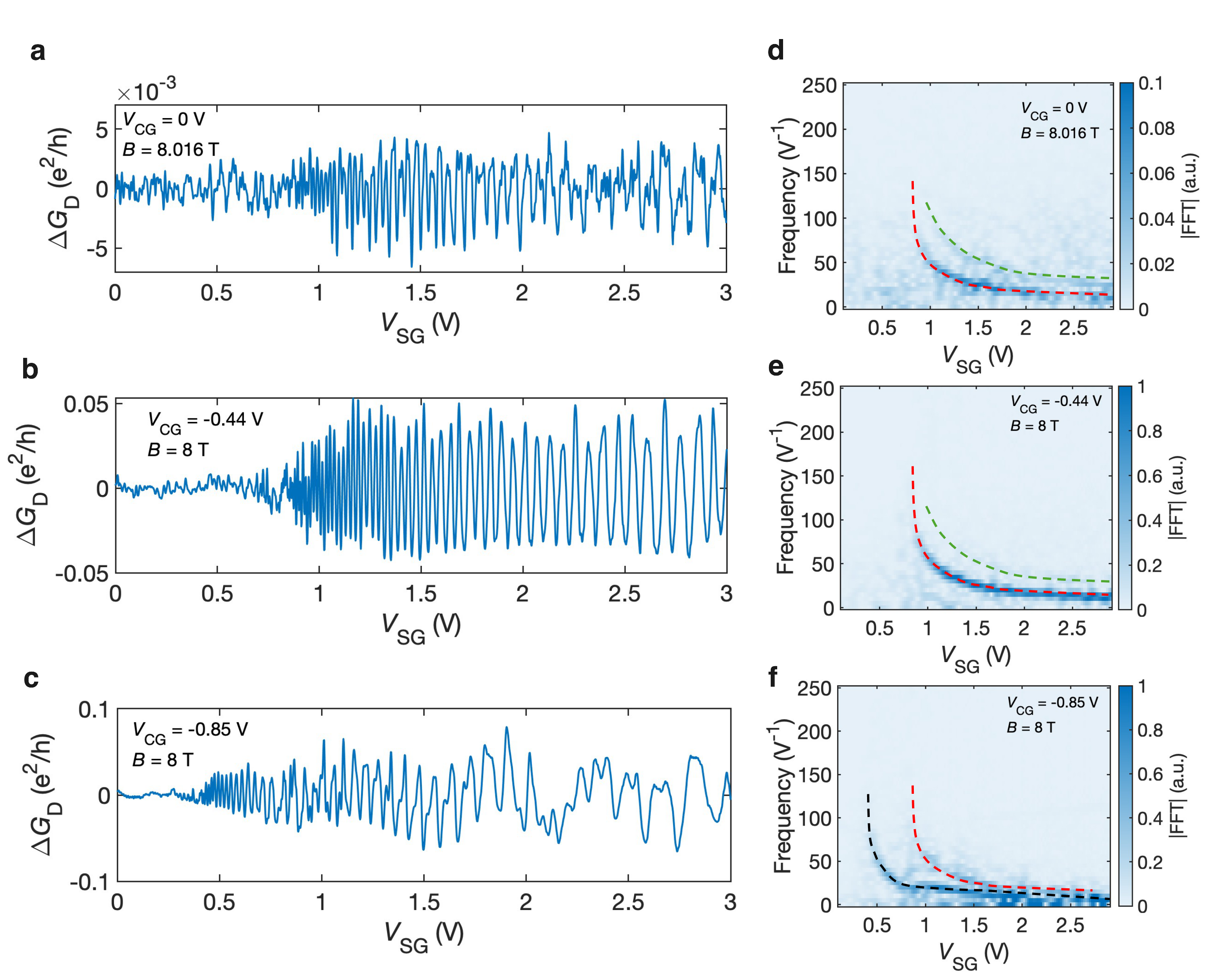}
    \caption{\textbf{Gate spectroscopy of the coupled edges of $\nu$ = -2 measured at 15 mK and at a magnetic field of around 8 T.} \textbf{a-c,} Single scans of diagonal conductance $\Delta G_{D}$ as a function of side gate voltage over a larger range of $V_{SG}$. The oscillations are measured at different $V_{CG}$ values, which correspond to different $\nu_{cavity}$ values, as shown in Fig.~\ref{fig:pyama_plots_8t_nu2}. \textbf{d-f,} Fourier transforms of \textbf{a-c} computed over a sliding window of $V_{SG}$. The red dashed lines in \textbf{d-f} represent the oscillations of the sum of both outer and inner edges. The green dashed lines in \textbf{d} and \textbf{f} indicate the second harmonic signals, which diverge at the same $V_{CG}$ value as the red dashed lines. The black dashed line in \textbf{f} represents the oscillations associated with the inner edges. Compared to Fig.~\ref{fig:pyama_plots_8t_nu2}a and d (measured at 250 mK), the edge coupling is stronger in Fig.~\ref{fig:shotgate_sweep_8t_nu2}a and d (measured at 15 mK). All the data are measured at zero bias voltage.}
    \label{fig:shotgate_sweep_8t_nu2}
\end{figure*}

First, we investigate the evolution of the oscillations as a function of the central gate voltage $V_{CG}$. Figure~\ref{fig:pyama_plots_8t_nu2}a-c provides results of the oscillations of the outermost edge at $\nu$ = -2, as a function of both $V_{SG}$ and magnetic field. Figure~\ref{fig:pyama_plots_8t_nu2}a shows the pyjama plot at $V_{CG}$ = 0, which means the filling factor 
is $\nu$ =-2 throughout the device. The positive slope of the oscillation indicates that the interference of the outermost edge is in the Aharonov-Bohm dominated regime. From the magnetic periodicity in Fig.~\ref{fig:pyama_plots_8t_nu2}d, we extract an interference area of 0.37~$\mu m^{2}$, which is in good agreement with the designed cavity size. 

Then we adjust the filling factor inside the cavity $\nu_{cavity}$ by varying $V_{CG}$, while keeping the filling factors in the surrounding regions unchanged. The relation between $V_{CG}$ and $\nu_{cavity}$ is obtained from the Hall conductance as a function of the top gate voltage at a fixed bottom gate voltage $V_{bg}$ = -0.63 V, which can be found in Fig.~\ref{fig:supp1S3}. For positive $V_{CG}$, corresponding to a reduced carrier density inside the cavity, a transition\cite{sivan_observation_2016,kim_aharonovbohm_2024} from the AB regime to the Coulomb-dominated (CD) regime can be observed as shown in Fig.~\ref{fig:supp1S1}a-c. In this regime, the reduced carrier density enhances charging effects and the interferometer behaves as a quantum dot. When applying negative $V_{CG}$ ($|\nu_{cavity}|$ $\geq$ -2), the carrier density inside the cavity is increased and additional edge channels are introduced into the cavity. As shown in Fig.~\ref{fig:pyama_plots_8t_nu2}b-c and Fig.~\ref{fig:supp1S1}c-i, the periodicity in the pyjama plot is reduced approximately by a factor of two, and the corresponding FFT results in Fig.~\ref{fig:pyama_plots_8t_nu2}d-f also reveal a transition from $h/e$ periodicity to a dominant $h/2e$ periodicity, demonstrating a doubling of the interference frequency. This behavior correlates with the presence of multiple edge channels inside the cavity, where the interactions between neighboring channels become more relevant. 

In addition to this doubling of frequency, we observed another frequency branch emerging in the Fourier spectra, as shown in Fig.~\ref{fig:pyama_plots_8t_nu2}e-f and Fig.~\ref{fig:supp1S2}f-i. This re-emergent component has the same magnetic field periodicity $1/\Delta B$ as the fundamental oscillation of the interfering outermost edge channel, but it becomes increasingly dominant under a more negative $V_{CG}$. To examine the origin of this re-emergent frequency, we measured the oscillations of the outermost channel at different bulk filling factors $\nu$ = -1 and -3 at different $V_{CG}$, which are provided in the Supplementary Materials. When $\nu$ = -3, the additional frequency branch appears before $V_{CG}$ = -0.46 V, corresponding to $\nu_{cavity}$ = -4. In contrast, in the $\nu$ = -1 case, only one single frequency branch is observed in the range from $V_{CG}$ = 0 V to -1.22 V ($\nu_{cavity}$ = -1 to -4). These observations indicate that this re-emergent frequency is associated with the presence of multiple edge channels and is very sensitive to the edge couplings.

\section{Gate spectroscopy at different magnetic fields}\label{sec:gate-spectroscopy}

\begin{figure*}
    \includegraphics[width=0.95\textwidth]{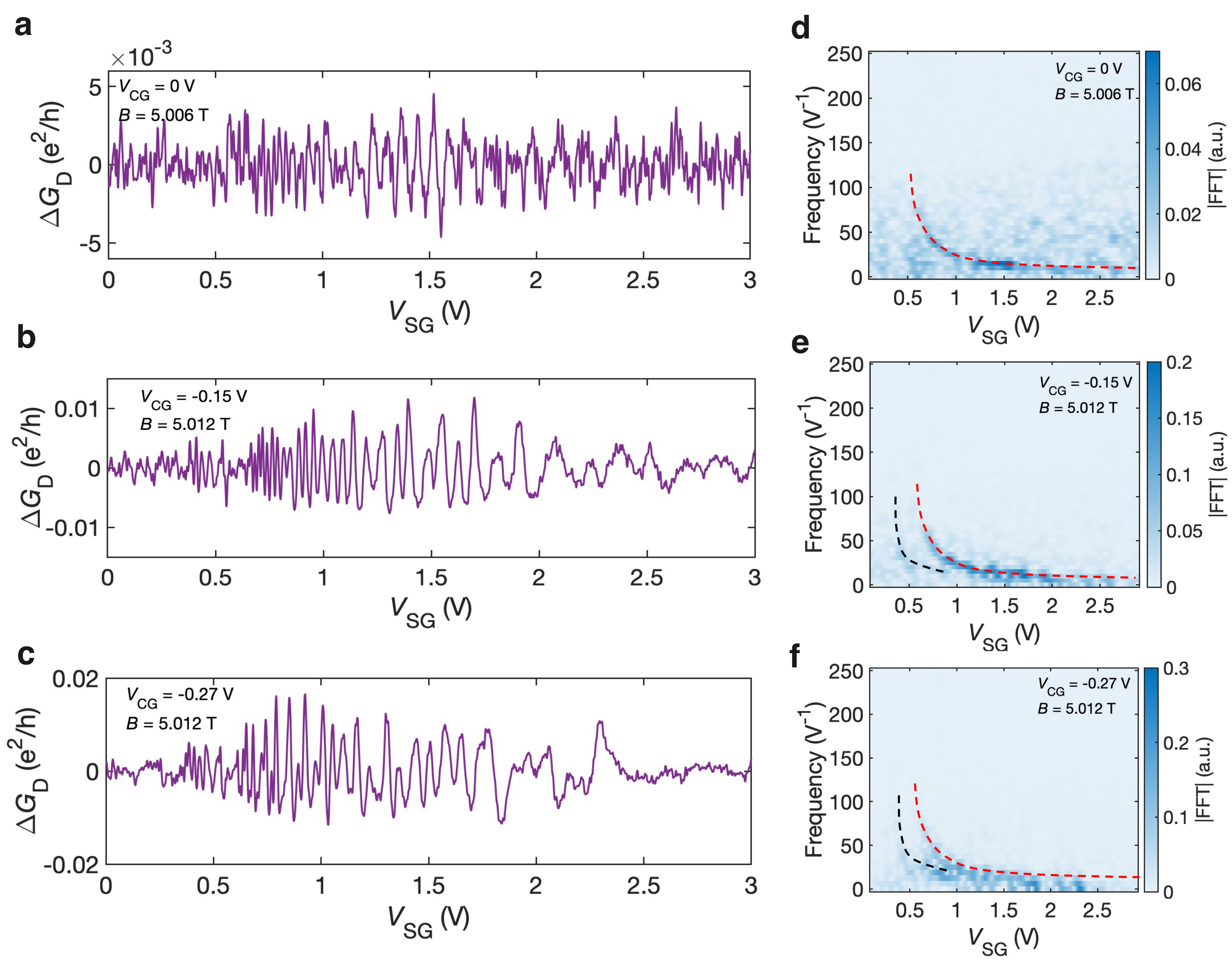}
    \caption{\textbf{Gate spectroscopy of the coupled edges of $\nu$ = -2 measured at 15 mK and at a magnetic field of around 5 T.} \textbf{a-c,} Single scans of diagonal conductance $\Delta G$ as a function of side gate voltage over a larger range of $V_{SG}$. The oscillations are measured at different $V_{CG}$ values, which correspond to different $\nu_{cavity}$ values, i.e., $V_{CG}$ = 0 V ($\nu_{cavity}$ = -2), $V_{CG}$ = -0.15 V ($\nu_{cavity}$ = -2.x) and $V_{CG}$ = -0.27 V ($\nu_{cavity}$ = -3), respectively. \textbf{d-f,} Fourier transforms of \textbf{a-c} computed over a sliding window of $V_{SG}$. The red dashed lines in \textbf{d-f} represent the oscillations of the sum of both outer and inner edges, and the black dashed lines in \textbf{e} and \textbf{f} represent the oscillations associated with the inner edges.}
    \label{fig:gate_spectroscopy_5t}
\end{figure*}

\begin{figure*}
    \includegraphics[width=0.95\textwidth]{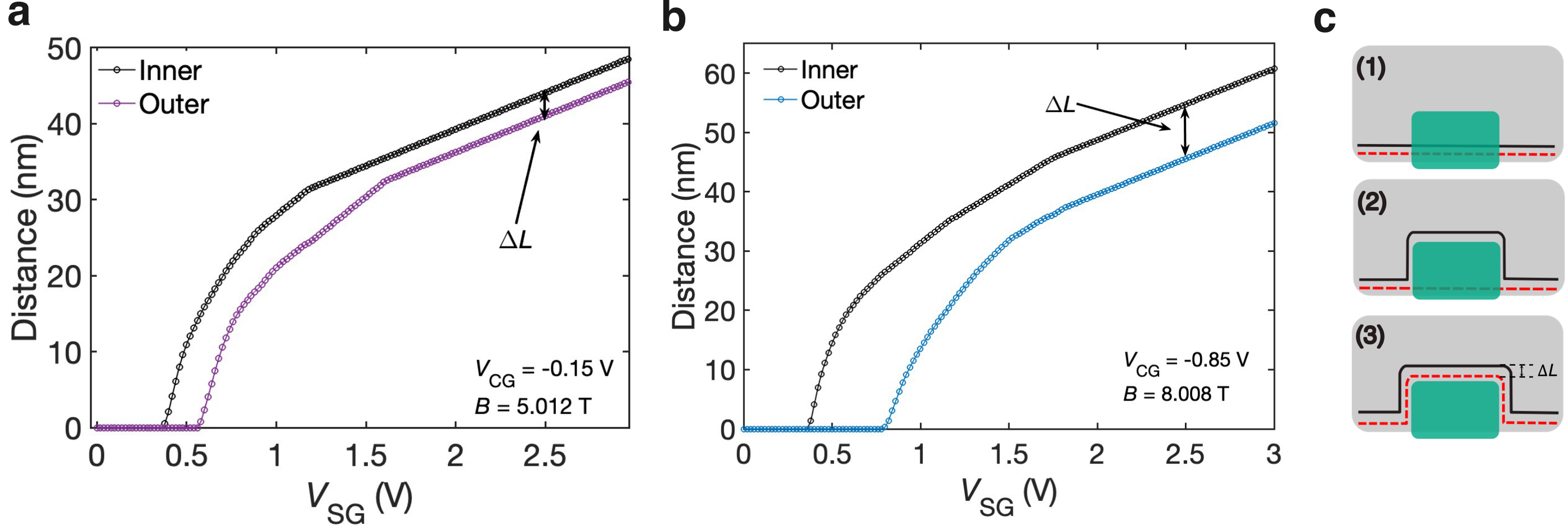}
    \caption{\textbf{Determination of the edge-to-side gate distances.} \textbf{a-b,} The edge-to-side gate distances as a function of $V_{SG}$ at different magnetic fields. The distances in \textbf{a} and \textbf{b} are obtained by integrating the $V_{SG}$ frequency dispersion shown in Fig.~\ref{fig:gate_spectroscopy_5t}e and Fig.~\ref{fig:shotgate_sweep_8t_nu2}f, respectively. The inner and outer edges are spatially separated by a distance $\Delta L$. \textbf{a,} At 5.012 T, the inner and outer channels are illustrated as black and purple lines, respectively. The distance between edges, $\Delta L$ $\approx$ 3 nm (at $V_{SG} = 2.5 V$), is less than the magnetic length $l_{B}$. In \textbf{b}, at 8.008 T, the distance ($\Delta L$ $\approx$ 9.2 nm) is comparable to $l_{B}$. Details of the edge-to-gate distance calculation are provided in the Methods. \textbf{c,} Illustrations of three different edge-channel configurations at $\nu$ =-2 when varying $V_{SG}$. The red and black lines correspond to the outer and inner edge channels, respectively. (1) When the filling factor beneath the side gate is smaller than -2, the two edge channels go directly through the region under the gate. (2) When the filling factor beneath the side gate is -2$<\nu<$-1, only the outer edge channel can go through the side gate region. (3) When $\nu>$-1, the two edge channels are expelled from the side gate region.}
    \label{fig:edge_distances}
\end{figure*}

To investigate the evolution of the inter-edge coupling, we measure $\Delta G_{D}$ oscillations as a function of $V_{SG}$ over an extended range of side gate voltages at different magnetic fields and different $V_{CG}$. Figure~\ref{fig:shotgate_sweep_8t_nu2} shows the results for different $V_{CG}$ measured at 15 mK and at a magnetic field of around 8 T. At zero $V_{CG}$, a pronounced frequency branch (shown as the red dashed line in Fig.~\ref{fig:shotgate_sweep_8t_nu2}d) occurs at $V_{SG}$ $\approx$ 1 V and gradually approaches a constant value at larger $V_{SG}$. A weak branch (green dashed line) close to the red dashed line follows a similar $V_{SG}$ dependence and emerges at approximately the same threshold voltage. We therefore attribute this feature to a higher harmonic of the dominant oscillation rather than to an independent edge channel. 

At a small $V_{CG}$ = -0.44 V, corresponding to $\nu_{cavity}$ ${\approx}$ -3, a single dominant frequency branch is present (shown as the red dashed line in Fig.~\ref{fig:shotgate_sweep_8t_nu2}e), corresponding to the coupled-edge interference regime shown in Fig.~\ref{fig:pyama_plots_8t_nu2}b and e. When decreasing $V_{CG}$ to -0.85 V, two distinct frequency branches emerge (shown in Fig.~\ref{fig:shotgate_sweep_8t_nu2}f as red and black dashed lines, respectively), consistent with the additional spectral component observed in Fig.~\ref{fig:pyama_plots_8t_nu2}f and Fig.~\ref{fig:supp1S2}. As mentioned in Sec.~\ref{sec:vcg-dependence}, the higher-frequency branch (red dashed line) evolves from the outer-edge interference observed at $V_{CG}$ =0 and develops into the coupled-edge interference when $V_{CG}$ decreases. In contrast, the additional frequency branch emerges at a smaller $V_{SG}$ value. Since the bulk filling factor is only controlled by the bottom gate and remains fixed at -2, the inner edge channel is expelled from beneath the side gate before the outer edge channel, which makes the inner edge channel respond at a lower $V_{SG}$ threshold. Therefore, we associate this additional frequency branch with the inner edge channel. Importantly, this inner edge channel is fully reflected at the QPCs, so that its appearance in the interference pattern provides a signature of coupling between the inner and outer edge channels.

We also repeat the gate-spectroscopy measurements at 5 T and 12 T, so that we can study the magnetic field dependence of the coupled-edge response. The gate-spectroscopy results at 5 T and 12 T are shown in Fig.~\ref{fig:gate_spectroscopy_5t} and Fig. S8, respectively. In the case of 5 T, two distinct frequency branches are resolved at $V_{CG}$ = -0.15 V, where the cavity filling factor is between $\nu_{cavity}$ = -2 and -3. In contrast, the second branch becomes more resolved only at a more negative $V_{CG}$, which corresponds to a more negative $\nu_{cavity}$. At 12 T (shown in the Supplementary Materials), only one single frequency branch is resolved over the gate voltage range. The earlier emergence of the second frequency branch with decreasing magnetic field indicates that the coupled-edge response is enhanced at lower magnetic fields.

Gate spectroscopy makes it possible to measure the spatial configurations of edge channels\cite{deprez_tunable_2021,yang_evidence_2024}, which influence the inter-edge coupling strength. To study the magnetic field dependence of the edge-coupling strength, we extract the distances between the inner and outer edge states at different magnetic fields as a function of $V_{SG}$ by integrating the frequency dispersions, i.e., those in Fig.~\ref{fig:shotgate_sweep_8t_nu2}f and Fig.~\ref{fig:gate_spectroscopy_5t}f. The details of the calculation can be found in the Methods. We provide the resulting positions of the edge channels at 5.012 T as a function of $V_{SG}$ in Fig.~\ref{fig:edge_distances}a and the results at 8.008 T in Fig.~\ref{fig:edge_distances}b. In the larger $V_{SG}$ regime (i.e., $V_{SG}$ $\geq$ 2.5 V), where high-visibility oscillations are found, one can compare the separations between the inner and outer edges at different magnetic fields. For instance, the distance between edges $\Delta L$ is approximately 3 nm at 5.012 T, which is substantially smaller than the magnetic length $l_{B}$ = 11.2 nm. In contrast, we found a larger $\Delta L$ $\approx$ 9.2 nm at 8.008 T, comparable to $l_{B}$ = 8.86 nm. The enhanced coupled-edge features at lower magnetic field coincide with a regime where the spatial separation of the two edge channels becomes smaller compared with the magnetic length, supporting the picture in which the spatial proximity enhances the inter-edge coupling.

\section{Discussion and conclusions}\label{sec:discussion}

We have investigated the Aharonov-Bohm interference of the outermost edge channel of bilayer graphene FPIs at $\nu$ = -2 using both electrostatic gates and magnetic fields to tune the configuration and coupling between neighboring quantum Hall edges. The electrostatic control of the cavity drives a transition from the conventional $h/e$ periodicity to a doubled interference. Besides, we observed a re-emergent frequency branch which is associated with the inner edge channel, despite this fully reflected channel seemingly not participating in the interference. The appearance of both spectral features is sensitive to the edge configuration, indicating a role of inter-edge coupling in determining the interference response. In particular, the inner edge feature is strengthened at large FPI $|\nu_\text{cavity}|$, and it only appears for bulk fillings $|\nu|\geq2$, when at least two hole Landau levels are filled in the bulk, whereas it is absent for $\nu=-1$ as can be seen in the Supplementary Materials, Fig.~\ref{fig:nu_1_pajamaplots} and Fig.~\ref{fig:nu_1_fft}. These findings can be naturally accommodated in a picture of period doubling caused by strong capacitive couplings between the edge modes, building on the theory of Ref.~\cite{werkmeister_strongly_2024}, which we turn to now.

We consider the FPI at $|\nu_\text{cavity}|\geq2$ and focus on the two outermost edge modes inside the cavity. If we denote the charge in the outermost interfering mode with $Q_1$ and the charge in the inner mode with $Q_2$, we can model the ground state energy with $E(\delta Q_1,\delta Q_2)=K_1\delta Q_1^2+K_2\delta Q_2^2+2K_{12}\delta Q_1\delta Q_2$, where $\delta Q_{1,2}=Q_{1,2}-Q_{1,2}^{(0)}$. Here $Q_{1,2}^{(0)}$ denotes the electrostatic equilibrium charge we would obtain without quantization conditions on the charges, and stability requires $K_{12}^2\leq K_1K_2$. As the outer mode is interfering, we allow $Q_1$ to take any real value, while $Q_2$ must be quantized to an integer, meaning that $\delta Q_2$ can only jump by integer values.

\begin{figure}
    \centering
    \includegraphics[width=0.95\linewidth]{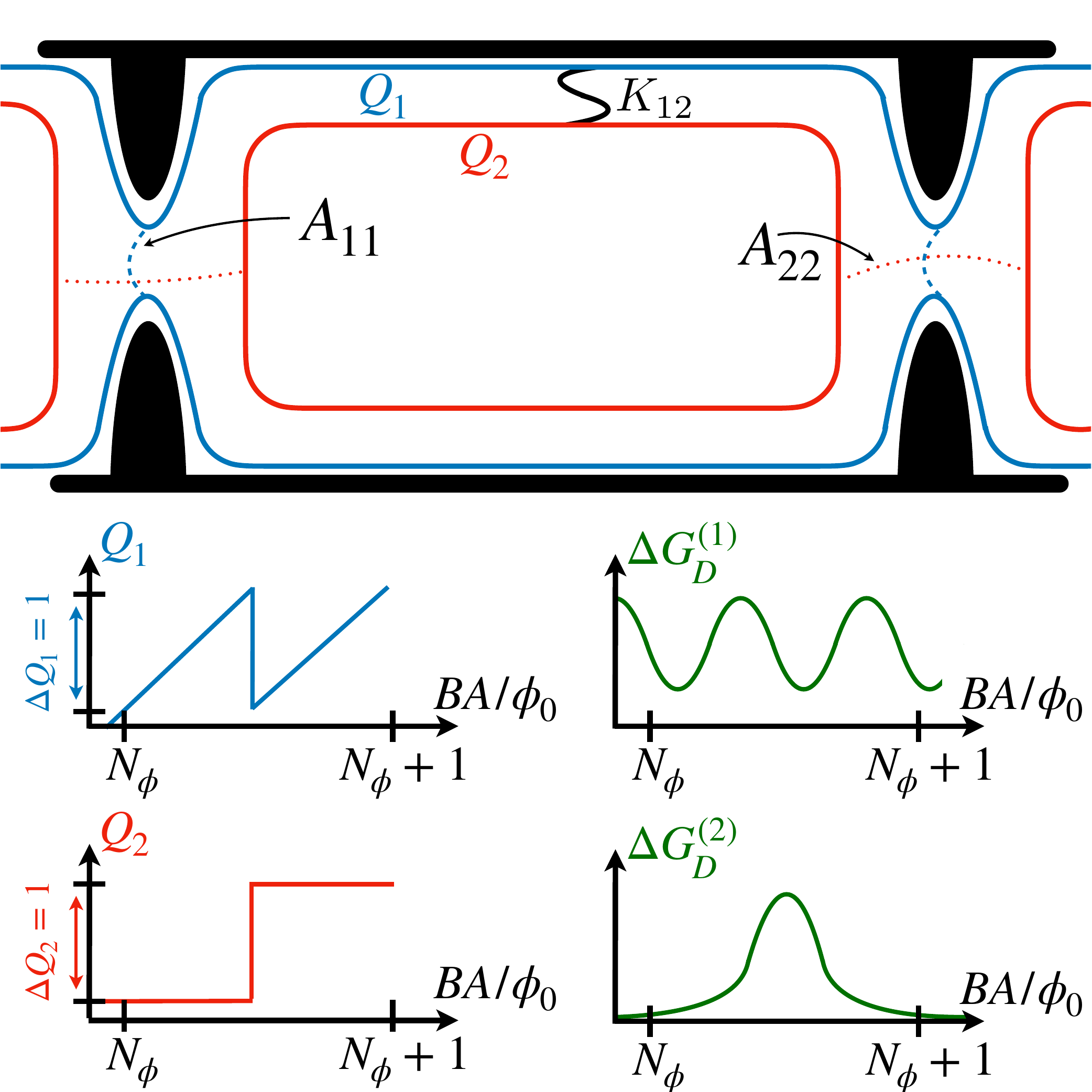}
    \caption{Top: A sketch of the edge modes in the FPI, showing the interfering outer mode (blue, $Q_1$) and the inner mode (red, $Q_2$). At the QPCs, the outer mode tunnels vertically via a term $A_{11}$, but weak tunneling of the inner mode in the horizontal direction ($A_{22}$) is also possible. Bottom: Due to the strong capacitive coupling $K_{12}$, $Q_1$ continuously climbs by one unit twice for each unit of flux added to the interfering area, leading to the double-period oscillation in $\Delta G_D^{(1)}$ of magnitude $|A_{11}|^2$. During the cycle, a charge is transferred between $Q_1$ and $Q_2$, and at exactly this point resonant tunneling due to $A_{22}$ gives a peak in $G^{(2)}_D$. The total response $\Delta G_D = \Delta G_D^{(1)}+ \Delta G_D^{(2)}$ reproduces both frequencies seen in Fig.~\ref{fig:pyama_plots_8t_nu2}f.}
    \label{fig:doubling_inner_sketch}
\end{figure}

In this interpretation, varying $B$ and $V_\text{SG}$ changes $Q_{1,2}^{(0)}$ continuously. The outer mode charge $Q_1$ adapts continuously, giving a signal $\Delta G_D^{(1)}\sim |A_{11}|^2\cos(2\pi Q_1+\delta)$ \cite{werkmeister_strongly_2024,Feldman_2022}, and $Q_2$ may jump by an integer value, which induces a jump $-K_{12}/K_{1}$ in $Q_1$. Whenever $K_{12}/K_1\approx 1$, this jump does not leave a signature in $\Delta G_D^{(1)}$. When a quantum of flux is threaded through the interferometer (by an appropriate increase in area or magnetic field strength), this leads to $Q_{1,2}^{(0)}\rightarrow Q_{1,2}^{(0)}+1$. The discontinuous jump by $\sim -1$, however, means that $Q_1$ must continuously increase by unity twice in this period, leading to two interference maxima and a doubling in the observed frequency, as sketched in Fig.~\ref{fig:doubling_inner_sketch}. A version of this is indeed seen for bulk filling $\nu=-1$, but curiously, bulk fillings $\nu\leq -2$ show another interference peak at a lower frequency at high $|\nu_\text{cavity}|$ (Fig.~\ref{fig:pyama_plots_8t_nu2}f), which can be interpreted as arising from the inner edge mode (Fig.~\ref{fig:shotgate_sweep_8t_nu2}c,f).

To reconcile this inner mode interference with the above picture which requires a quantized $Q_2$, we augment the model of an FPI for bulk $\nu\leq -2$ with weak tunneling between the $n=-2$ IQHE edge mode outside the interferometer and the (nearly) fully reflected $n=-2$ mode inside the FPI, as shown in Fig.~\ref{fig:doubling_inner_sketch}. Denoting by $\tilde E(Q_2)=\min_{Q_1} E(Q_1,Q_2)$ the equilibrium energy for a given inner mode charge $Q_2$, we observe that this tunneling process will be at resonance whenever $\tilde E(Q_2)\approx\tilde E(Q_2+1)$, producing a signature $\Delta G_D^{(2)}\sim |A_{22}|^2$ whenever this condition holds. For each additional flux quantum added to the interfering area, this condition is satisfied exactly once, in particular at the point coinciding with the (invisible) phase slip in $Q_1$ and jump in $Q_2$. At this point, we expect a signal from the inner mode; see Fig.~\ref{fig:doubling_inner_sketch}. The doubled-period outer mode $\Delta G_D^{(1)}$ and the inner mode $\Delta G_D^{(2)}$ oscillations thus originate from different physical tunneling processes, yet their periods are determined by the same energetic considerations of capacitive couplings between the edge modes, with $\Delta G^{(1)}_D$ oscillating at roughly double the frequency of $\Delta G^{(2)}_D$.

The tunneling term $|A_{22}|^2$ leading to $G^{(2)}_D$ can only exist for $\nu\leq -2$, and is expected to be strongest for large $|\nu_{cavity}|$ where edge modes are pushed further towards the edges of the FPI, both of which are indeed seen in experiment. Within our model, the relative strengths of the two peaks (inner and double-frequency) should depend sensitively on the quantum point contact voltages $V_\text{QPC}$; further constricting the quantum point contacts should strengthen the ``vertical'' tunneling in Fig.~\ref{fig:doubling_inner_sketch} and thus the interference of the outer edge mode and the doubled frequency, while weakening the ``horizontal'' tunneling and thus reducing the prominence of the peak associated with the inner frequency. Opening the QPCs would have the reverse effect on both peak heights. Probes of this dependence would be a beneficial direction for future work and could offer further insights into how the edge modes interact.

Furthermore, the magnetic field dependence provides further insight into the coupled-edge regime. Gate spectroscopy reveals that the additional frequency branch emerges earlier when the magnetic field is reduced. By computing the edge trajectories from the gate-dependent frequency dispersion, we find that the separation between the two edges is substantially smaller than the magnetic length at 5.012 T, while at 8.008 T the separation increases to approximately 9.2 nm, which is comparable to the corresponding magnetic length. This observation coincides with the enhanced coupled-edge spectral features in the gate-spectroscopy results and supports the picture in which the spatial proximity of neighboring edges can play a vital role in the emergence of the anomalous interference spectrum.

It is worth considering the charging effects when the edge-bulk coupling is not negligible\cite{rosenow_influence_2007,halperin_theory_2011,sivan_observation_2016}, which can also modify the interference patterns. In the limit of strong edge-bulk coupling, the FPIs are in the Coulomb-dominated regime\cite{rosenow_influence_2007,zhang_distinct_2009,halperin_theory_2011,sivan_observation_2016}, where the oscillations are independent of the magnetic field. Although the transition from the CD to the AB regime can be accessed in our device by tuning $V_{CG}$, the measurements discussed in the main text are performed in the AB regime, suggesting the inter-edge coupling is dominant.

Moreover, our results represent a systematic study of tuning the inter-edge coupling through both electrostatic gating and magnetic field, and provide a route towards controlling the interactions between quantum Hall edge states and understanding the origin of anomalous interference frequencies.

\section*{Methods}\label{sec:methods}
\subsection*{Sample preparation}\label{sec:sample-preparation}
The hBN-encapsulated bilayer graphene heterostructure is stacked using standard dry transfer methods and then transferred onto 285 nm thick highly doped SiO$_{2}$ chips. The top graphite is divided into several regions by e-beam lithography and etched using an oxygen plasma etching process. The air bridge is fabricated on top of the top graphite by using a three-layer resist process (PMMA 950 A4/MMA/PMMA 495 A4), followed by a deposition process of Cr/Au (25 nm/250 nm).

\vspace{1em}

\subsection*{Determination of edge-to-gate distances}\label{sec:edge-distance-method}
The determination of edge-to-gate distances follows the calculation mentioned in the literature\cite{deprez_tunable_2021,yang_evidence_2024}. Denoting the distance between the side gate and the interfering edge by $\delta$, the lever arm of the side gate can be written as $\alpha(V_{SG}) = L_{SG}\frac{d\delta}{dV_{SG}}$, which represents the efficiency with which the side gate affects the edge channel. We can compute the distance as a function of $V_{SG}$ by integrating the gate-voltage dependence of the frequency of the corresponding edge as $\delta(V_{SG}) = \frac{\Phi_{0}}{BL_{SG}}\int_{V_{C}}^{V_{SG}}f(V)$d$V$, where $\Phi_{0}$ is the magnetic flux quantum, $V_{C}$ is the side gate voltage when the frequency curves start to diverge, and $L_{SG}$ is the edge length of the side gate, which in our calculation is 0.43 $\mu$m. Here we also use the relation $\Delta A = \frac{\Phi_{0}}{B} = \alpha \Delta V_{SG} = \alpha/f$.

\vspace{1em}






\newpage

\bibliography{references}



\vspace{1em}
\noindent \textbf{Acknowledgements}

We thank Moty Heiblum, Hyungkook Choi and Benjamin Sacépé for their helpful discussions. We acknowledge the support of SNSF (Eccellenza Grant No. PCEGP2\_194528) and support from the QuantERA II Programme, which has received funding from the European Union’s Horizon 2020 research and innovation programme (Grant Agreement No. 101017733). K.W. and T.T. acknowledge support from the JSPS (KAKENHI Grant Nos. 20H00354 and 23H02052) and the World Premier International Research Center Initiative, MEXT, Japan. G.W. is supported by the SNSF (Ambizione Grant No. PZ00P2-216183). 

\vspace{1em}
\noindent \textbf{Author Contributions}

M.B. and N.M. conceived the project. N.M. made the stacks, fabricated the devices, and performed the measurements. N.M. analyzed the data with input from M.B. T.L. and S.H.S. performed the theoretical calculations. K.W. and T.T. provided the hBN crystals. N.M. and T.L. wrote the manuscript with input from all the authors.

\vspace{1em}
\noindent \textbf{Competing Interests}

\noindent The authors declare that they have no competing interests.

\newpage




\clearpage

\onecolumngrid
\setcounter{equation}{0} 
\setcounter{figure}{0} 
\setcounter{table}{0} %
\renewcommand{\thefigure}{S\arabic{figure}}
\begin{figure*}
    \centering
    \includegraphics[width=1\linewidth]{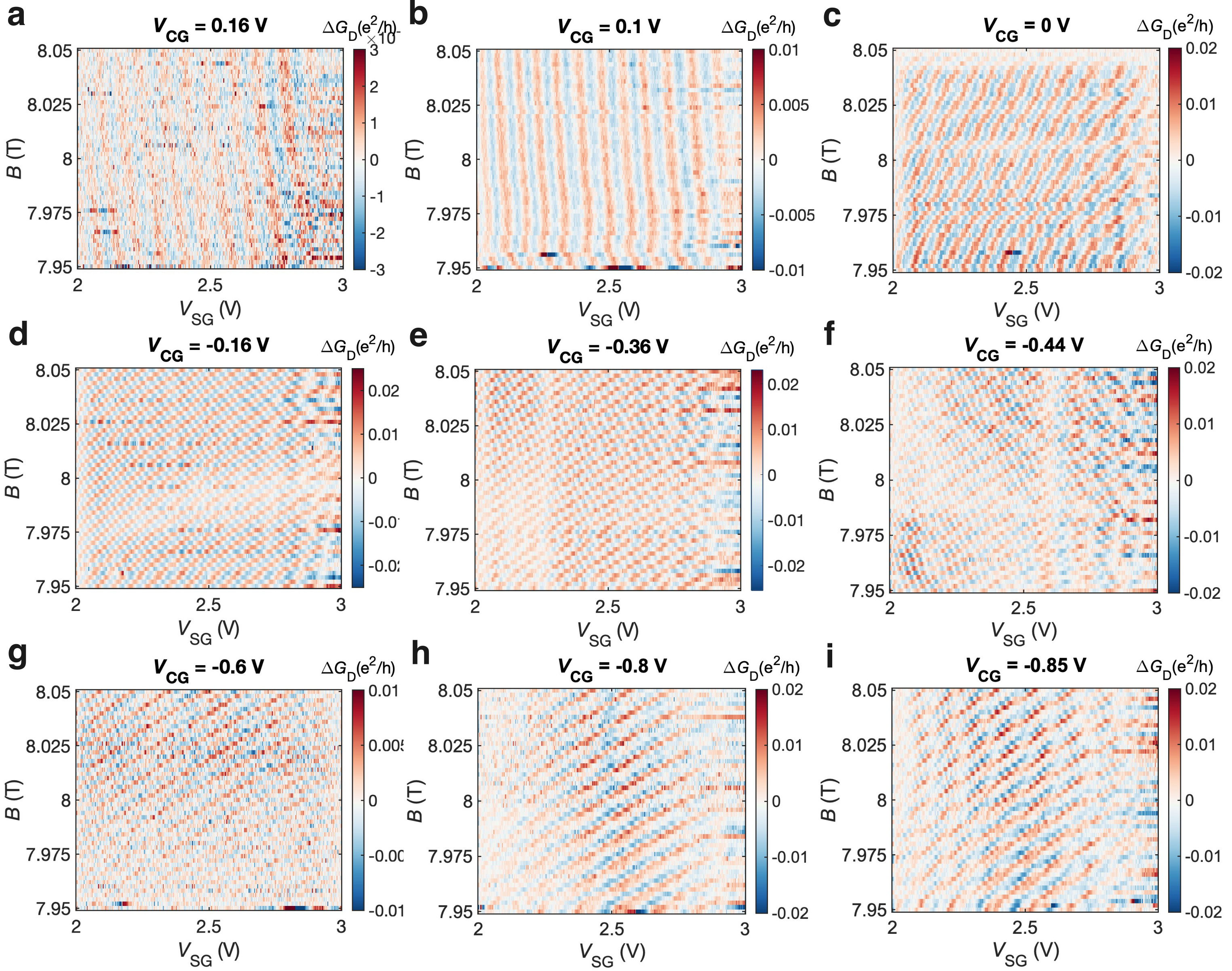}
    \caption{\textbf{Pyjama Plots of the outer edge channel of $\nu$ = -2 at different $V_{CG}$.} First of all, a transition from Coulomb-dominated regime to Aharonov-Bohm dominated regime are shown from \textbf{a} to \textbf{c} when decreasing $V_{CG}$ from 0.16 V to zero. A transition from $h/e$ to $h/2e$ frequency occurs when applying more negative $V_{CG}$, i.e., from 0 to -0.16 V. When decreasing $V_{CG}$ to -0.6 V, i.e., \textbf{g} to \textbf{i}, phase jumps are observed at $V_{CG}$ =-0.85 V in \textbf{i}. All the data are obtained at 250 mK.}
    \label{fig:supp1S1}
\end{figure*}

\begin{figure*}
    \centering
    \includegraphics[width=1\linewidth]{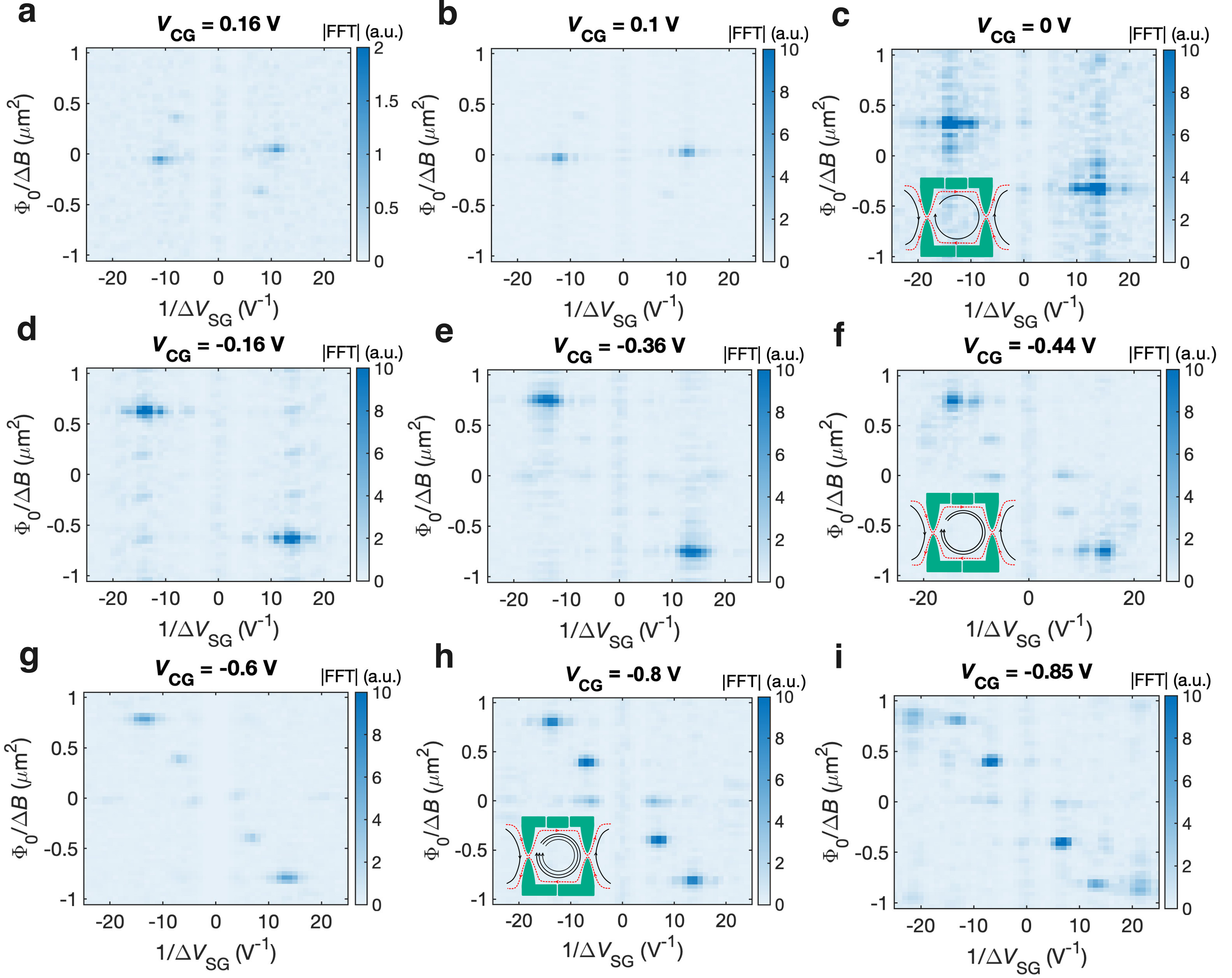}
    \caption{\textbf{2D FFT of the $\Delta G_{D}$ oscillations shown in Fig.S1.} The inner edge channel of $\nu=-2$ is fully reflected and only the outermost edge mode travels through QPCs. The periodicity of magnetic field in \textbf{c} indicates the designed FP area is around 0.4 $\mu m^{2}$. Both $B$ and $V_{SG}$ periodicities are increasing when more negative $V_{CG}$ values are applied, as shown in \textbf{c-f}. Another lower frequency peak starts to emerge from $V_{CG}$ = -0.44 V, as shown in \textbf{f}, and it becomes more dominated when more channels are introduced by tuning $V_{CG}$, as shown in \textbf{f-i}. The inset schematics in \textbf{c}, \textbf{f} and \textbf{h} show the edge channel configurations at different $V_{CG}$, in which the outermost channel is shown as red dahsed line and the inner edges are shown in black lines, same as the schematics shown the main text in Fig.2.}
    \label{fig:supp1S2}
\end{figure*}

\begin{figure*}
    \centering
    \includegraphics[width=0.5\linewidth]{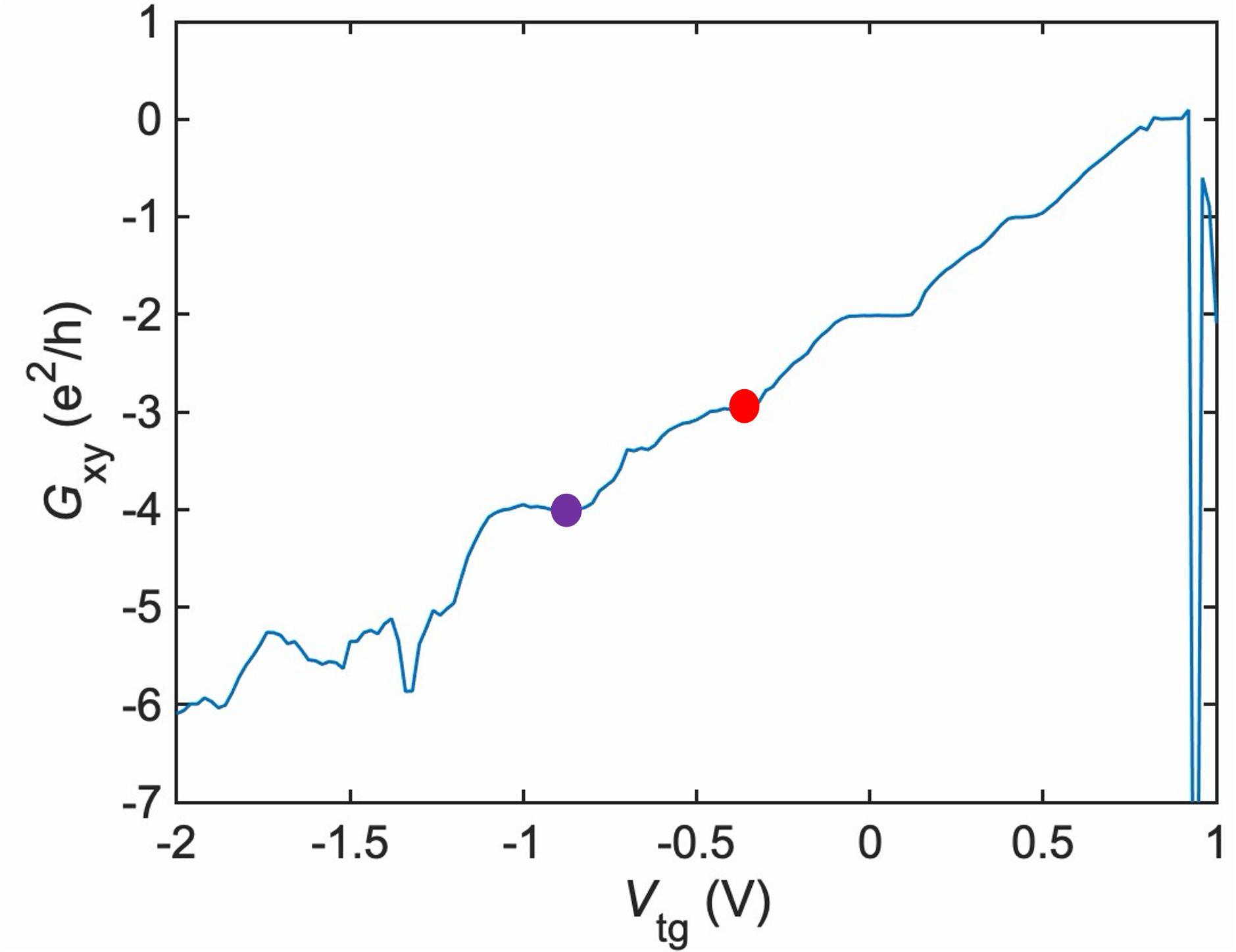}
    \caption{\textbf{Hall conductance as a function of $V_{tg}$ at $V_{bg}$ = -0.63 V and $B$ = 8 T.} The purple and red dots are marked as the setpoints of the voltage values of $V_{CG}$ of Fig. 2c and Fig. 2b in the main text, which correspond to $\nu_{cavity}$= -4 and -3, respectively. The measurement is performed at 250 mK.}
    \label{fig:supp1S3}
\end{figure*}

\begin{figure*}
    \centering
    \includegraphics[width=1\linewidth]{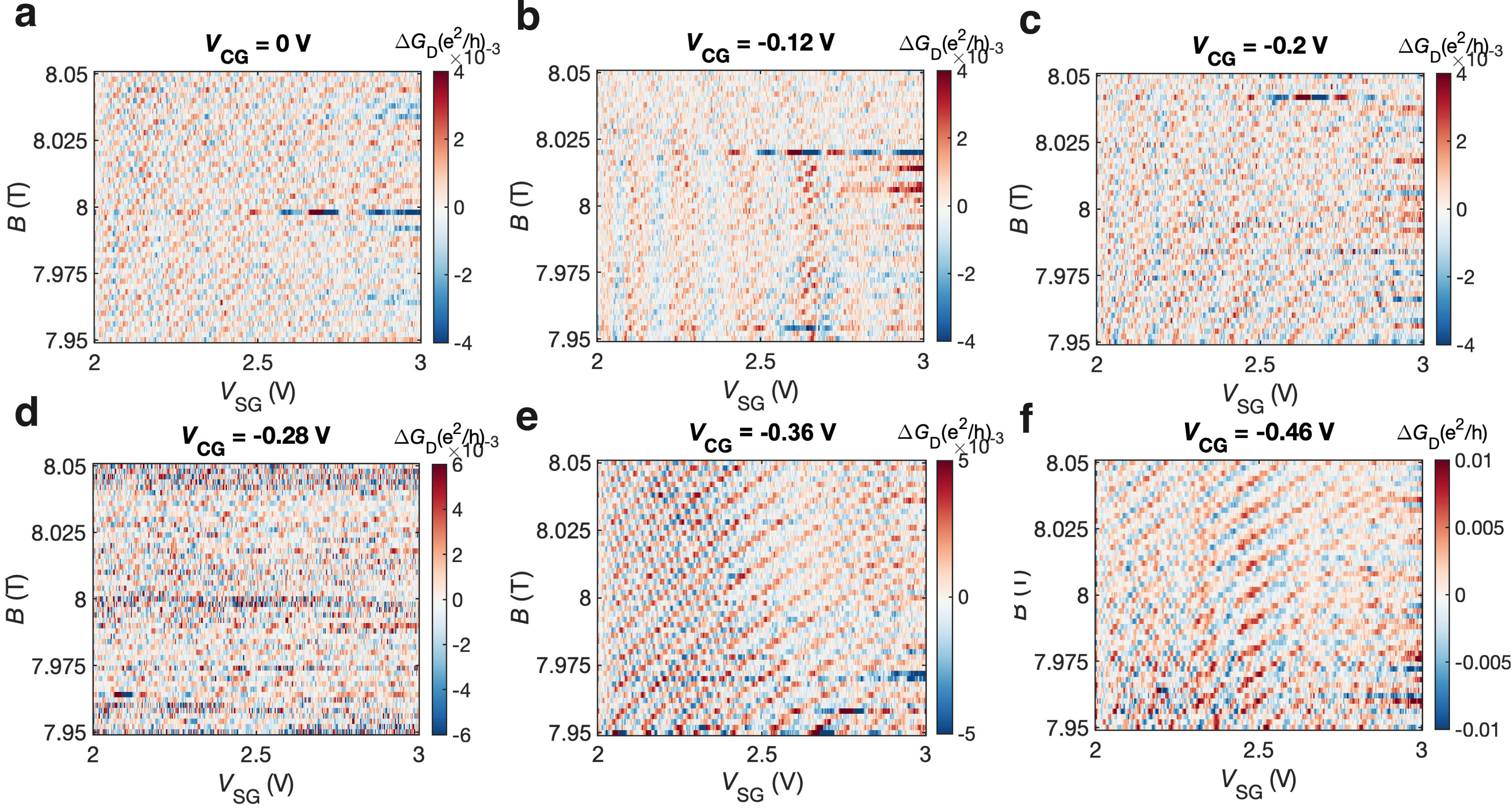}
    \caption{\textbf{Pyjama Plots of the outermost edge channel of $\nu$ = -3 at different $V_{CG}$.} From $V_{CG}$ = 0 V, the periodicity of the magnetic fields indicates a doubling behavior similar to the case shown in Fig. S1. When decreasing $V_{CG}$ to -0.2 V, as shown in from \textbf{c} to \textbf{f}, another frequency start to emerge. All the data are obtained at 250 mK.}
    \label{fig:supp1S4}
\end{figure*}

\begin{figure*}
    \centering
    \includegraphics[width=1\linewidth]{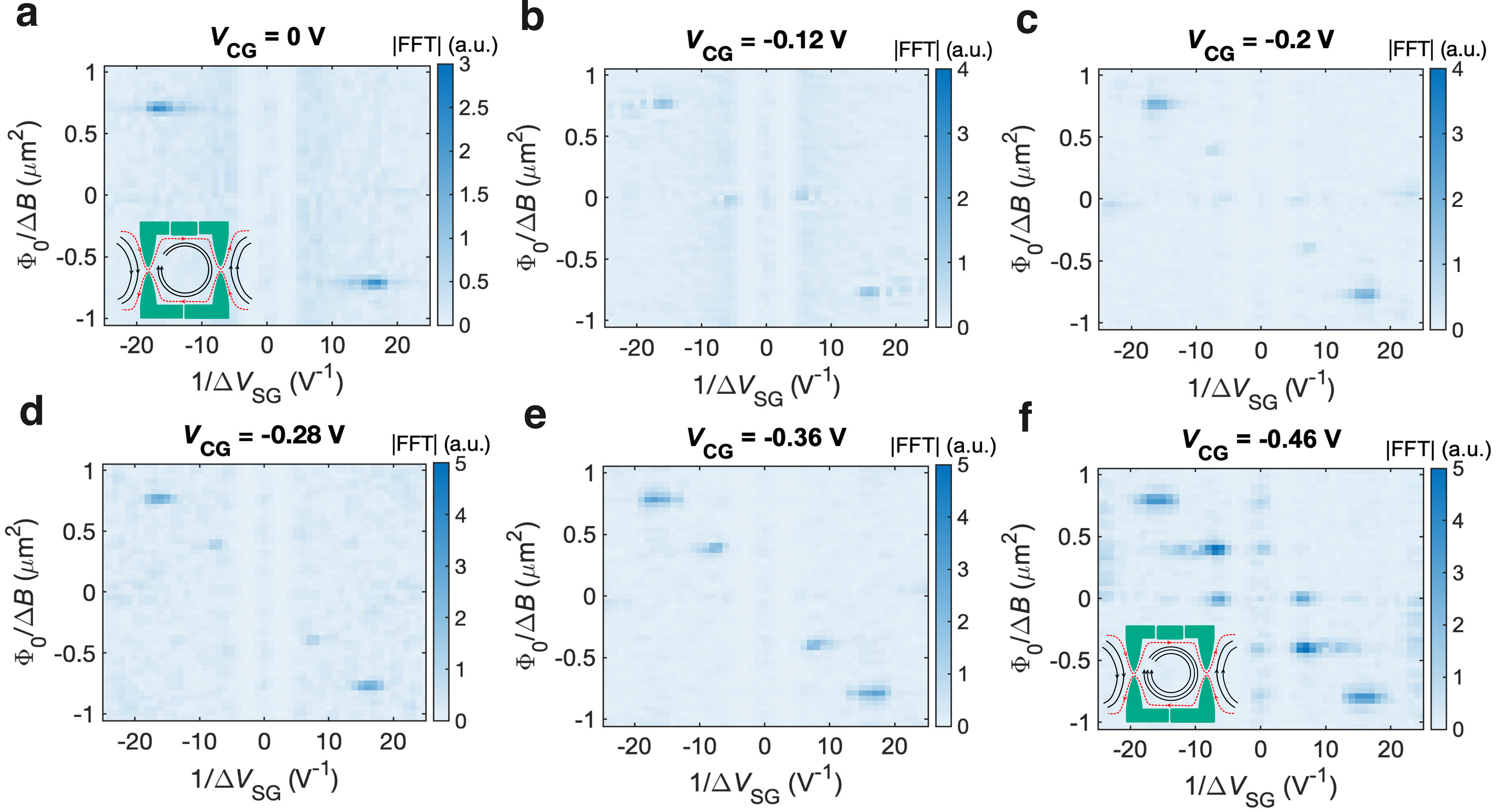}
    \caption{\textbf{2D FFT of the $\Delta G_{D}$ oscillations in Fig. S4.} Only the outermost edge channel of $\nu=-3$ travels through the QPCs. The doubling of periodicity in magnetic field is observed from \textbf{a} to \textbf{f}. A lower frequency peak occurs from $V_{CG}$ = -0.2 V, shown in \textbf{c}. The inset schematics in \textbf{a} and \textbf{f} indicate the edge channel configurations at $\nu_{cavity}$ = -3 and -4, respectively. The outermost edge channel is shown as the red dashed lines and all the other edge channels are plot as black lines.}
    \label{fig:supp1S5}
\end{figure*}

\begin{figure*}
    \centering
    \includegraphics[width=1\linewidth]{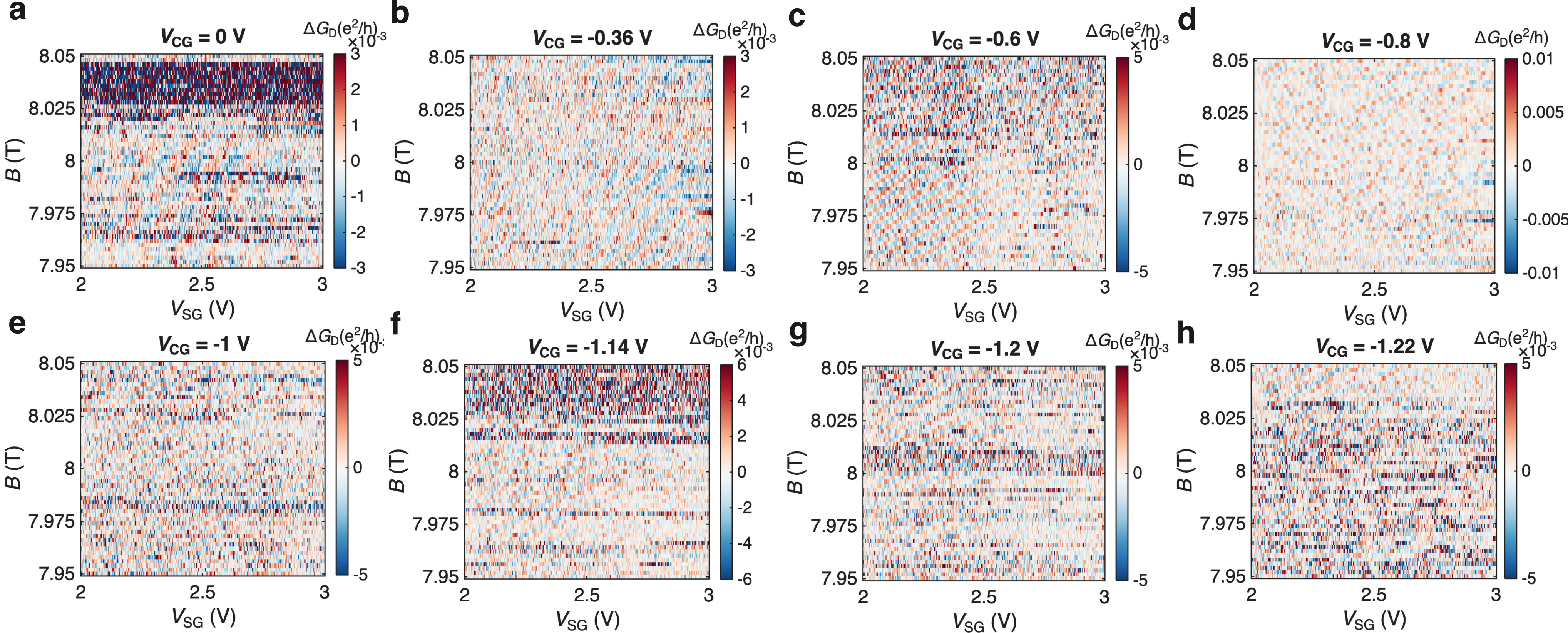}
    \caption{\textbf{Pyjama Plots of the channel of $\nu$ = -1 at different $V_{CG}$.} The $V_{CG}$ varies from 0 to -1.22 V, corresponding to a range of $\nu_{cavity}$ from 0 to -4. The periodicity of magnetic field and side gate is gradually increasing from \textbf{b} to \textbf{h}, indicating a transition from single edge oscillation to edge-coupling oscillation. All the data is obtained at 250 mK.}
    \label{fig:nu_1_pajamaplots}
\end{figure*}

\begin{figure*}
    \centering
    \includegraphics[width=1\linewidth]{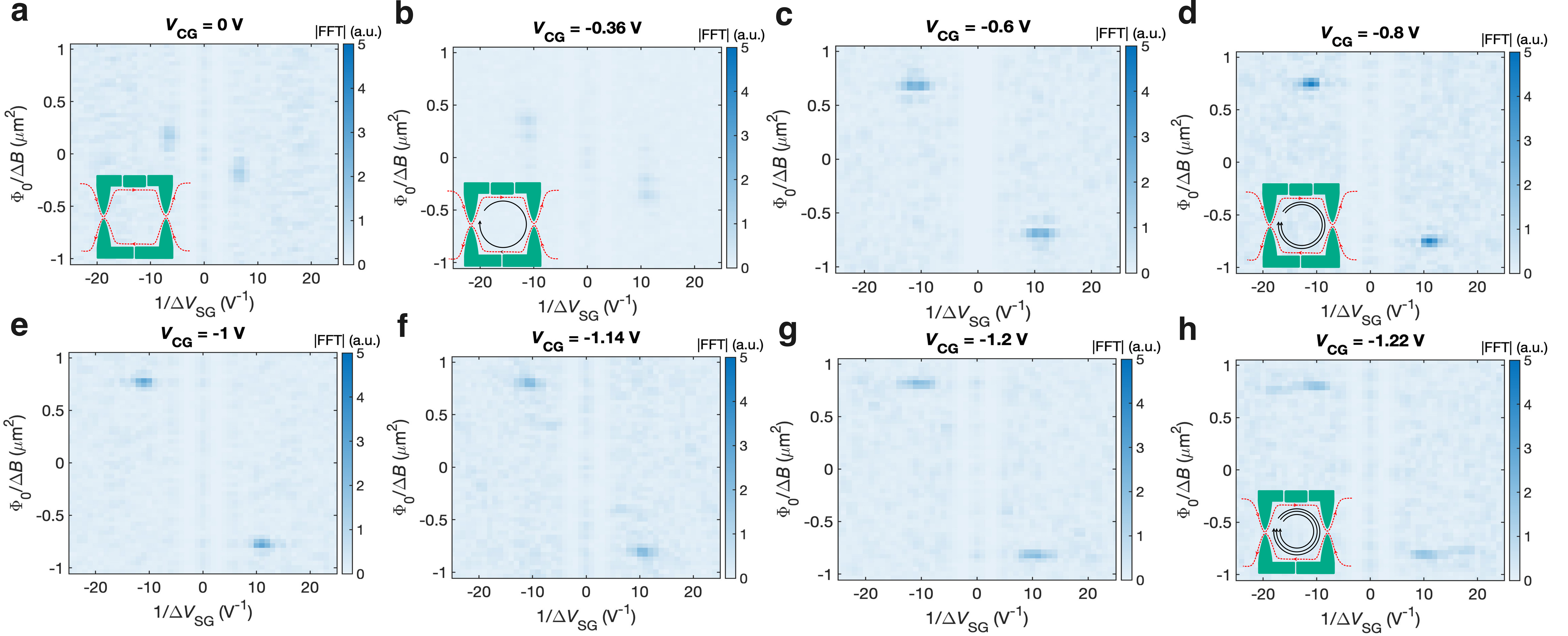}
    \caption{\textbf{2D FFT of the $\Delta G_{D}$ oscillations in Fig. S6.} Unlike the cases of $\nu=-2$ and $\nu=-3$, shown in Fig. S2 and Fig. S5, respectively, only one single frequency in all the $\nu_{cavity}$ values in the measurement of $\nu=-1$. This observation is consistent with the picture that the newly observed frequency shown in Fig. S2 and Fig. S5 is associated with inner edge that tunnels into the cavity. The inset schematics in \textbf{a}, \textbf{b}, \textbf{d} and \textbf{h} indicate the edge channel configurations at $\nu_{cavity}$ = -1, -2, -3 and -4, respectively. The outermost edge channel is shown as red dashed lines and all the other edge channels are shown as black lines.}
    \label{fig:nu_1_fft}
\end{figure*}

\begin{figure}
    \centering
    \includegraphics[width=0.8\linewidth]{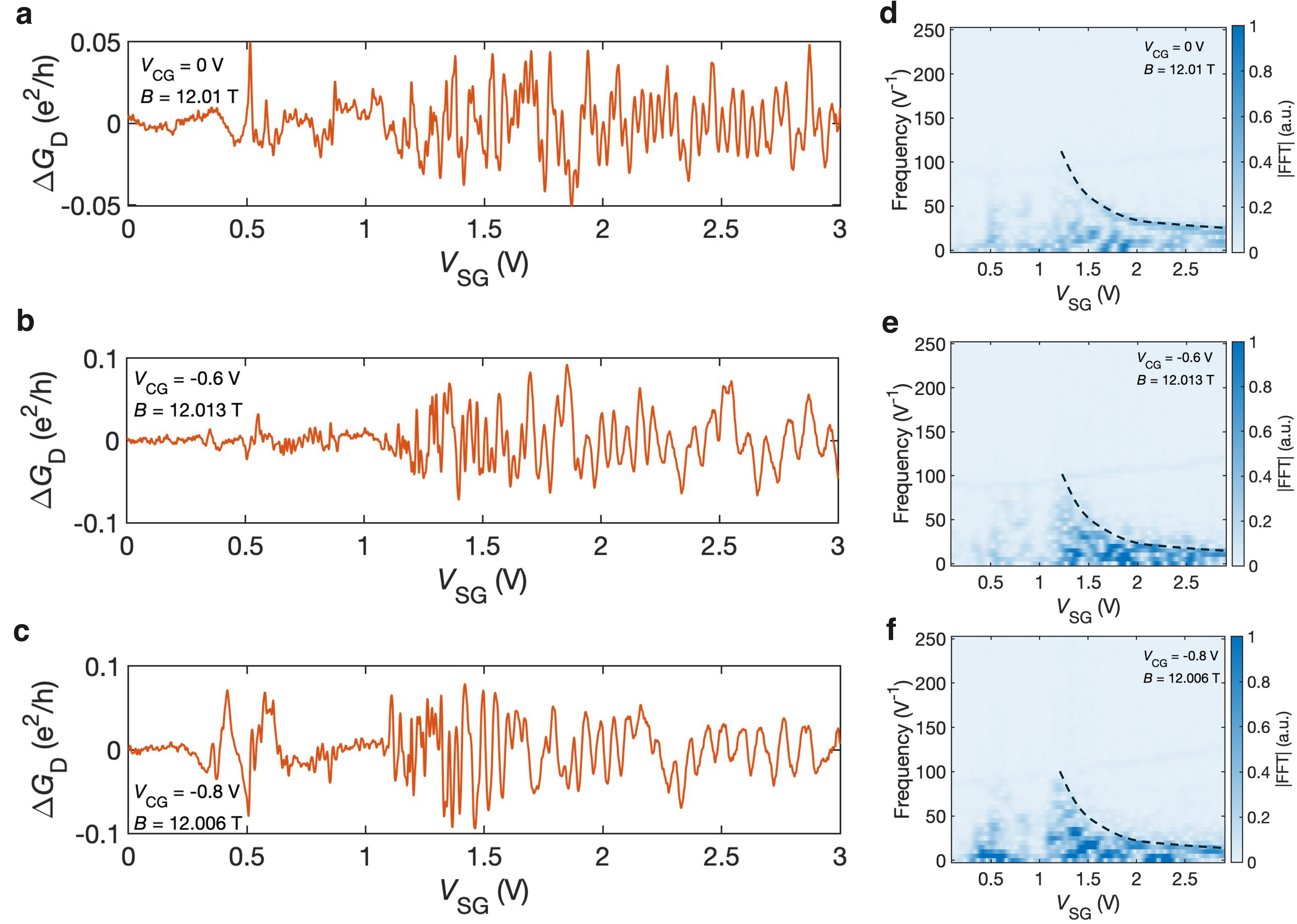}
    \caption{\textbf{Gate spectroscopy of the outermost channel of $\nu$ = -2 measured at 15 mK and at a magnetic field of around 12 T.} \textbf{a-c,} Single scan of diagonal conductance as a function of side gate in a larger range of $V_{SG}$. The oscillations are measured at different $V_{CG}$ values. \textbf{d-f,} Fourier transforms of \textbf{a-c} computed over a sliding window of $V_{SG}$. The black dashed lines in \textbf{d-f} represent the oscillations of the outermost edge. Unlike the cases of 5 T and 8 T, the Fourier results of 12 T show only one single frequency branch in all the $V_{CG}$ cases.}
    \label{fig:supp1S8}
\end{figure}

\end{document}